# Deformation mechanisms of Mg-Ca-Zn alloys studied by means of micropillar compression tests

Jingya Wang[a], Yiwen Chen[a], Zhe Chen[b], Javier Llorca[c, d, *], Xiaoqin Zeng[a, b, **]

[a] National Engineering Research Center of Light Alloy Net Forming,
Shanghai Jiao Tong University, Shanghai 200240, PR China.

[b] State Key Laboratory of Metal Matrix Composites,
Shanghai Jiao Tong University, Shanghai, 200240, PR China.

[c] IMDEA Materials Institute, 28906 Getafe, Madrid, Spain.

[d] Department of Materials Science, Polytechnic University of Madrid,
E. T. S. de Ingenieros de Caminos, 28040 Madrid, Spain.

*Corresponding author. E-mail address: javier.llorca@imdea.org.
**Corresponding author. E-mail address: xqzeng@sjtu.edu.cn.

**Abstract:**

The effect of Ca and Zn in solid solution on the critical resolved shear stress (CRSS) of <a> basal slip, tensile twinning and <c+a> pyramidal slip in Mg alloys has been measured through compression tests on single crystal micropillars with different orientations. The solute atoms increased the CRSS for basal slip to ~ 13.5 MPa, while the CRSS for pyramidal slip was lower than 85 MPa, reducing significantly the plastic anisotropy in comparison with pure Mg. Moreover, the CRSSs for twin nucleation and growth were very similar (~ 37 MPa) and the large value of the CRSS for twin growth hindered the growth of twins during thermo-mechanical processing. Finally, evidence of <a> prismatic slip and cross-slip between basal and prismatic dislocations was found. It is concluded that the reduction of plastic anisotropy, the activation of different slip systems and cross-slip and the weak basal texture promoted by the large CRSS for twin growth are responsible for the improved ductility and formability of Mg-Ca-Zn alloys.

# 1. Introduction

Mg is lightest among the structural metals and Mg alloys are currently considered for applications in transport (automotive and aerospace) and health care [1–3] due to their high specific stiffness, superior biocompatibility and biodegradability, ease of recycling and abundant reserves. However, Mg alloys generally exhibit low ductility and poor formability that originate from the low-symmetry hexagonal closed packed (hcp) lattice structure and the large differences in the critical resolved shear stress (CRSS) for dislocation glide between basal and non-basal slip [4–6]. As a result, plastic deformation of Mg alloys is dominated by basal <a> dislocations and {10-12} tensile twinning, which are the softest deformation modes, the latter to accommodate deformation along the $c$ axis direction. Twinning is a polar mechanism that is only activated when the deformation tends to increase the length of the $c$ axis, leading to a very large plastic anisotropy in the mechanical response of textured alloys, which triggers fracture during forming operations [7].

It is expected that Mg alloys with higher ductility and formability can be attained if additional pathways for dislocation motion can be introduced. This can be achieved if the ratio of the CRSS between <c+a> pyramidal and <a> basal slip is reduced and/or cross-slip is facilitated [8–11]. Moreover, microstructures with weak basal texture present reduced the plastic anisotropy during deformation along different orientations, enhancing formability [12]. These objectives can be pursued through the addition of the appropriate alloying elements. It has been widely reported that the texture of Mg alloys containing Ca [13], Ce [14] and Y [12] is much more random than that found in typical Mg-Al [15], Mg-Zn [16] and Mg-Al-Zn alloys [7,17]. Furthermore, experimental investigations and theoretical simulations have indicated that the presence of Li [18], Ca [16,19,20], Ce [21], Y [10,22,23] and Nd [8] in solid solution improves the ductility of Mg alloys by the activation of other deformation mechanisms on top of basal slip and tensile twining. For instance, Sabat et al. [21] found that the enhanced ductility of Mg-0.2Ce (wt.%) alloys originated from the simultaneous contribution of basal, prismatic and pyramidal <c+a> slip. Sandlöbes et al. [23] indicated that the improved ductility of Mg-Y alloys was related to the activation of <c+a> pyramidal dislocations. Curtin et al. [10] reported that solute-accelerated cross-slip contributed to the ductility Mg-Y alloys, and this prediction was confirmed by transmission electron microscopy (TEM) observations [24].

Large improvements in ductility have also been reported in Mg-Ca and Mg-Ca-Zn alloys [16,19,20,25–28] which present the additional benefits of reduced cost and improved biocompatibility, as compared with Mg alloys containing rare earths. First principles calculations and molecular dynamics simulations pointed out that Ca atoms in solid solution promote the activation of prismatic and pyramidal II slip as well as basal-to-prismatic cross-slip [29–31]. Furthermore, Yuasa et al. [32] also predicted that the co-existence of Zn and Ca atoms would facilitate basal-to-prismatic cross-slip. These predictions were in agreement with a number of experimental observations. For instance, Zhu et al. [20] found that the enhanced ductility (tensile elongation > 15%) of an extruded Mg-0.47Ca (wt.%) binary alloy could be traced to a large density of pyramidal I <a> dislocations during deformation. Zeng et al. [16] observed extensive evidence of pyramidal II <c+a> slip traces in a deformed Mg-0.3Zn-0.1Ca (wt.%) sheet, resulting in a tensile elongation over 20%. Similarly, Wang et al. [19] concluded -as a result of *in situ* electron backscatter diffraction (EBSD) and viscoplastic self-consistent simulations- that massive prismatic slip was the crucial factor for the high ductility (> 30%) of Mg-1.8Zn-0.2Ca (wt.%) alloy. In addition, Kang et al. [25] reported that the activation of <c+a> pyramidal dislocations was necessary to enhance the formability of extruded Mg-2Zn-0.2Ca (wt.%) alloy.

These experimental results were carried out in polycrystalline samples and it was difficult to determine the actual values of the CRSS for each slip system and twinning because of the influence of different contributions (texture, grain size, cross-slip) to the overall mechanical response. However, novel micromechanical testing techniques have been recently used to assess the fundamental deformation mechanisms of single crystals of Mg [24,33–42], Ti [43,44], Al [45,46], and Ta [47] alloys as well as of intermetallic compounds [48,49]. In particular, micropillar compression tests have been used to measure the CRSS for different slip systems as well as CRSS for twin nucleation and growth in pure Mg [33–37] and different Mg alloys [24,38–42] at ambient and elevated temperature as a function of the solute or precipitate content. This methodology is applied in this investigation to elucidate the intrinsic effect of solid solution atoms on the deformation mechanisms of Mg-Ca-Zn alloys. *In situ* and *ex situ* micropillar compression tests on single crystals with different orientations were carried out. The stress-strain curves were analyzed in combination with scanning electron microscopy (SEM), TEM and transmission Kikuchi diffraction (TKD) to ascertain the influence of Ca and Zn on the CRSS for slip and twinning. This information sheds light on the origins of the high ductility of Mg-Ca-Zn alloys and indicates potential avenues to improve the ductility and formability of Mg alloys.

## 2. Material and experimental techniques

A Mg-0.2Ca-1.8Zn (wt.%) alloy was casted from commercially pure Mg and Zn as well as a Mg-20 wt.% Ca master alloy in an electric resistance furnace under a protective atmosphere of $CO_2$ (99 vol. %) and $SF_6$ (1 vol. %) gas mixture. The actual chemical composition of the cast ingots was analyzed using Inductively Coupled Plasma Atomic Emission Spectroscopy (ICP-AES). Afterwards, the ingots were solution treated at 400°C for 12h, followed by extrusion at 250°C with an extrusion ration of ~18:1. More details about the processing steps can be found in Ref. [19]. Small discs of 10 $mm^3$ were cut from the extruded samples and homogenized within quartz capsules filled with Ar at 450 °C for 6 days to promote grain growth. The disk surface was mechanically ground using abrasive SiC papers until a grit size of 7000 and chemically etched with a solution of 75 ml ethylene glycol, 24 ml of distilled water and 1 ml of nitric acid to remove the residual surface damage and to reveal the grain boundaries.

The crystallographic orientation of the grains in the discs was obtained by EBSD using a Tescan Mira-3 SEM equipped with an Oxford Instruments Nordlys EBSD detector at an acceleration voltage of 20 kV and step size of 10 μm. The EBSD data were analyzed with Channel 5 software and MTEX Toolbox v5.2.8 [50]. Grains with specific orientations to favor particular deformation modes were selected. Micropillars with square cross-section of 5 x 5 $μm^2$ and an aspect ratio in the range 2:1 to 3:1 were milled in the center of the selected grains using a Tescan Lyra-3 Focused Ion Beam (FIB) - SEM dual beam microscope with a Ga+ ion beam operated at 30 kV. The initial beam current was 15 nA to remove the material quickly and was reduced to 100 pA in the final polishing step to minimize surface damage. Moreover, the sample was tilted an extra 0.5° with respect to the ion beam axis during the final milling step to fabricate micropillars with a tapper angle < 1°.

Micropillar compression tests were carried out *ex situ* (Hysitron Triboindenter TI950 system) and *in situ* (Hysitron PI88 system) within the SEM. It was reported previously that size effects on the flow stress for different deformation modes (basal and pyramidal slip as well as twinning) were minimum in Mg and Mg alloys for micropillars of these dimensions [41,42]. A diamond cylindrical flat punch of 10 μm in diameter was attached to the indenter tip to compress the micropillars. All the tests were carried out at a constant strain rate of $10^{-3}$ s$^{-1}$ under displacement control up to a maximum strain of 10%. The experimental displacement, $D_{ex}$, was corrected to account for the elastic deformation of the material beneath the micropillar and of the diamond indenter according to [51,52]:

$$D = D_{ex} - \frac{1-\nu_i^2}{E_i}\left(\frac{P}{d}\right) - \frac{1-\nu_b^2}{E_b}\left(\frac{P}{d}\right) \quad (1)$$

where $P$ is the load and $d$ is the pillar diameter. $E_i$ and $\nu_i$ stand for the elastic modulus (1140 GPa) and Poisson's ratio (0.07) of diamond, while $E_b$ and $\nu_b$ stand for the elastic modulus and Poisson' ratio (0.35) of the Mg-Ca-Zn alloy. The elastic modulus was determined for each orientation by means of nanoindentation tests with a Berkovich tip in the same grains where the micropillar compression tests were performed, following the methodology presented in [51]. Nevertheless, the differences in the elastic moduli between different orientations (Table 1) were very small because the elastic anisotropy of Mg is very reduced. The engineering stress-strain curves were obtained from the load-corrected displacement data using the cross-sectional area and the height of the micropillar before deformation.

The morphology of the undeformed and deformed micropillars was characterized using secondary electrons in the SEM (Tescan Mira-3). Then, thin foils parallel to the micropillar axis were extracted from the compressed micropillars using FIB. The deformation mechanisms were analyzed by Transmission Kikuchi Diffraction (TKD) and the dislocation structures were observed through transmission electron microscopy (TEM). The TKD maps were collected at 20 kV with a step size of 70 nm in the Tescan Mira-3 microscope. The TEM observations were performed in a Talos F200X G2 microscope at an accelerating voltage of 200 kV. The dislocations in the foils were studied using the two-beam condition and the dislocation type was identified based on the dislocation extinction condition, i.e. dislocations are visible when $\vec{g} \times \vec{b} \neq 0$ ($\vec{g}$: diffraction vector and $\vec{b}$: Burgers vector).

## 3. Results
### 3.1 Microstructure and deformation modes

A representative EBSD map of the Mg-Ca-Zn alloy after the homogenization treatment is depicted in Fig. 1. The crystallographic orientations are randomly distributed because the co-existence of Ca and Zn weakens the texture [16]. Grains were large (> 100 µm) and the micropillars were carved in the center of the grains to ensure that they were single crystals. Grains with particular orientations that favored the activation of a specific slip or twin system during micropillar compression were selected, as indicated in Fig. 1. They are grain A (the loading direction of the micropillar forms an angle of ~50° with [0001]), grain B (loading direction close to [0001]), grain C (loading direction nearly parallel to [11-20]) and D (loading direction [10-10]). The angle between the *c* axis of each grain and the compression direction is listed in Table 1, together with the elastic modulus obtained by nanoindentation tests and the highest Schmid factors (SF) corresponding to basal slip <a>, tensile twinning, prismatic slip <a>, pyramidal I <a>, pyramidal I <c+a> and pyramidal II <c+a> slip. Basal slip presents the highest SF in grain A, which is deformed at ~50° from the [0001] orientation, and is expected to dominate plastic deformation. Grains B and C are suitably oriented to activate pyramidal II <c+a> slip during micropillar compression. Finally, grain D favors tensile twining that is expected to control the deformation as the SF for basal slip is very low (0.03).

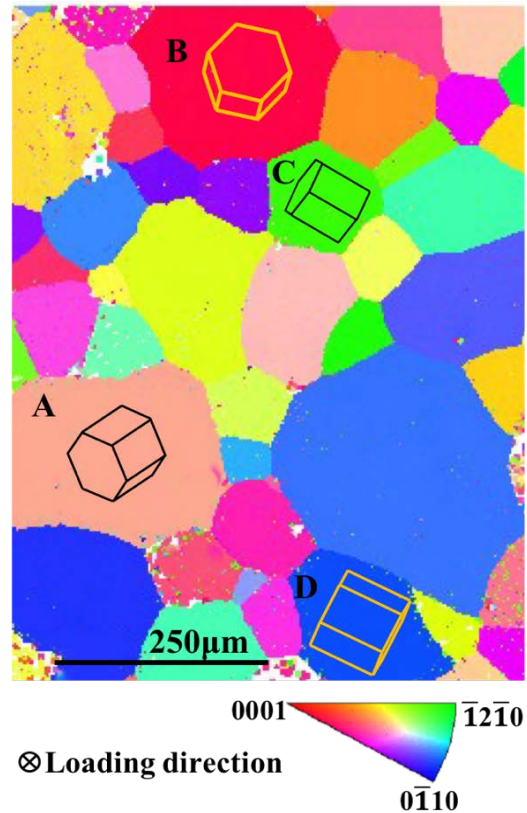

**Fig. 1**. EBSD map showing the orientations of the crystals (as given by inverse pole figure) in the Mg-Ca-Zn alloy after the homogenization treatment. The grains selected to carve the micropillars are denoted by A (loading direction forms an angle of ~50° with respect to [0001]), B (close to ([0001]), C (nearly parallel to [11-20]) and D (along [10-10]), shown by the orientation of the hcp lattice. The compression loading direction was perpendicular to the image.

**Table 1.** Inclination angle of the micropillar axis with respect to [0001], elastic modulus parallel to the loading direction, and highest Schmid Factors of the main deformation modes with respect to the loading direction of the Mg-Ca-Zn alloy micropillars.

| Loading direction | Inclination angle (°) | Elastic modulus (GPa) | Highest Schmid Factors | | | | | |
|---|---|---|---|---|---|---|---|---|
| | | | Basal slip | Tensile twin | Prismatic slip | Pyramidal I <a> | Pyramidal I <a+c> | Pyramidal II <c+a> |
| A [25-7-6] | 50 | 36.4 | 0.48 | 0.19 | 0.29 | 0.41 | 0.35 | 0.43 |
| B [0001] | 17 | 34.5 | 0.23 | -* | 0.03 | 0.14 | 0.48 | 0.49 |
| C [11-20] | 78 | 36.7# | 0.21 | 0.35 | 0.42 | 0.42 | 0.45 | 0.49 |
| D [11-10] | 88 | 36.7 | 0.03 | 0.49 | 0.47 | 0.43 | 0.43 | 0.39 |

*: Tensile twinning cannot be activated during compression along the *c*-axis.

#: The elastic modulus of Grain C is assumed to be same as D because both of them are deformed along the *a*-axis.

### 3.2 Deformation mechanisms of micropillars in grain A

Representative engineering stress-strain curves of micropillars of grain A are plotted in Fig. 2a. The x axis was shifted by 1% for each curve for the sake of clarity. The curves show gradual yielding of the micropillar with the applied strain, that is followed by a region where deformation progresses at constant stress. Large strain bursts are observed in this plateau region and this behavior is similar to that observed in Mg, Mg-Al, Mg-Zn and Mg-Y micropillars deformed in equivalent orientations [24,33,36,42]. The yield stress (to determine the initial CRSS for plastic deformation) was taken from the critical points (denoted by black stars in Fig. 2a), where the linear portion of stress-strain curves (due to elastic deformation) deviates from linearity. The precise construction to determine the yield points is detailed in previous investigations [38,42].

Secondary electron SEM micrographs of one micropillar of grain A are shown before (Fig. 2b) and after compression (Figs. 2c and 2d), the latter from two adjacent lateral view-directions of the front side and the right side, as marked in Fig. 2b. Distinct slip traces appear in both lateral surfaces, but most of the deformation is localized along a few slip bands, which are associated with the strain bursts in the stress-strain curves. The morphology of the deformed micropillars is very similar to that reported in Mg, Mg-Al, Mg-Zn and Mg-Y micropillars suitably oriented for basal slip [24,33,36,42]. The orientation of the basal slip plane (indicated by the blue plane) and of the relevant slip directions (marked by the red arrows) were identified using VESTA [53] for both lateral surfaces of the micropillar and are plotted in Figs. 2e and 2f. It is evident that the slip traces are parallel to the basal plane and that shear takes place along the [11-20] direction. The (0001) <11-20> basal slip system has the highest SF during micropillar compression, as shown in Table 2. Thus, it can be concluded that micropillars from grain A deformed by basal slip and the CRSS for basal slip was 13±0.5 MPa, as calculated from the yield stress and the corresponding SF.

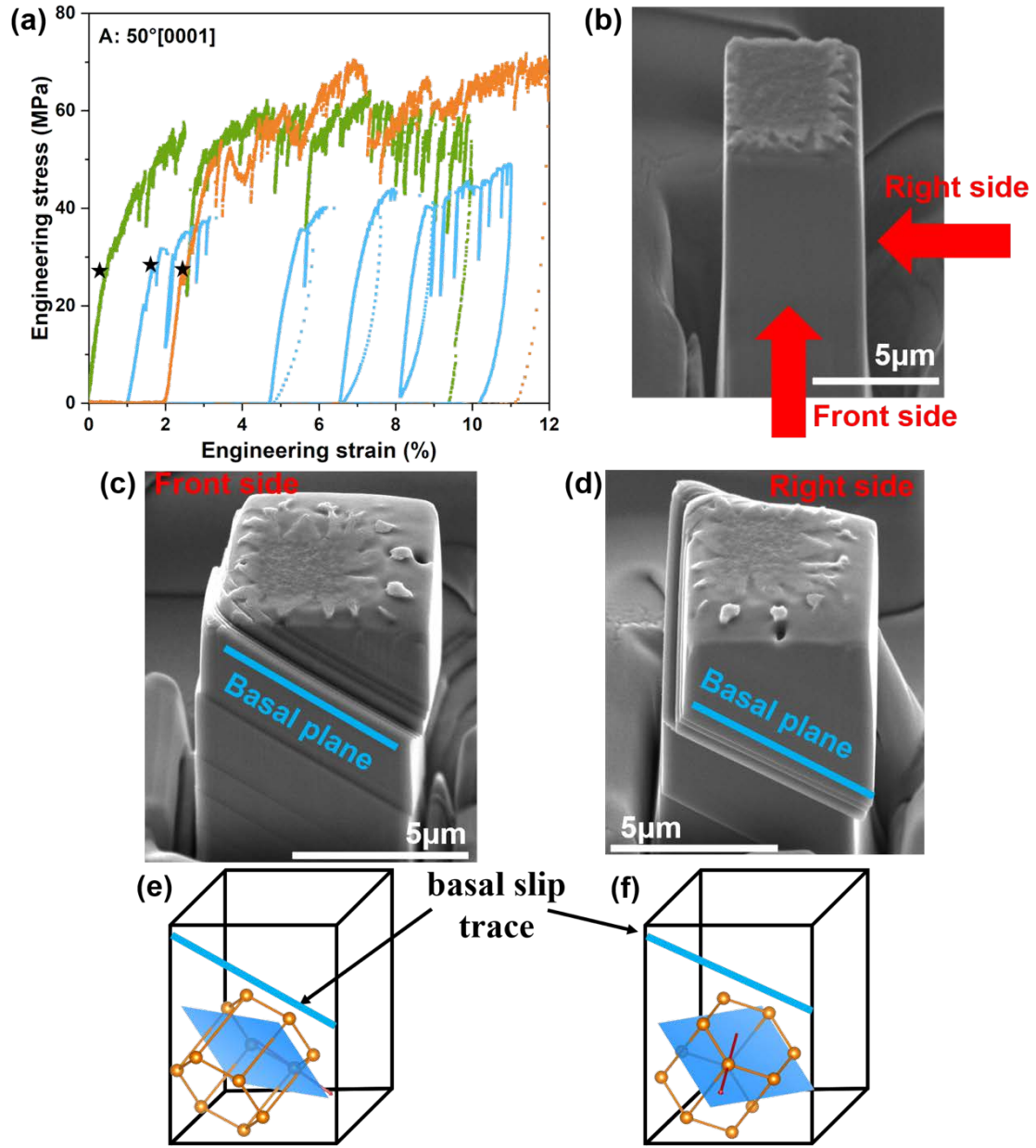

**Fig. 2.** (a) Engineering stress-strain curves of micropillar compression tests along direction A. The black stars denote the yield stress of each curve. (b) SEM micrograph of the micropillar before deformation and after compression observed from (c) front side and (d) right side. The corresponding schematics of the crystallographic lattice in the outline of the pillar are shown in (e) for the front side and (f) for the right side. The blue planes indicate the basal glide planes, and the red arrows represent the relevant shear directions.

### 3.3 Deformation mechanisms of micropillars in grain B (close to [0001])

The engineering stress-strain curves obtained from the compression tests of micropillars carved from grain B (parallel to [0001]) are depicted in Fig. 3a. After an initial elastic region, the yield stress that marks the onset of plastic deformation is easily identified by a "knee" in the stress-strain curve. Afterwards, the stress-strain curves show a marked strain hardening as well as continuous serrations of different intensity. Stress-strain curves with a marked yield point followed by linear hardening were also reported in micropillars deformed in compression along the *c* axis in Mg-4Al (at.%), 9Al (at.%) [41], Mg-0.4 (wt.%) and Mg-4Y (wt.%) [24] alloys, although the yield stress in all these cases was much higher (> 200 MPa). Moreover, serrations were also observed in the stress-strain curves in these investigations but they were less pronounced than those in Fig. 3a. On the contrary, micropillar compression tests along [0001] of Mg-4Zn (wt.%) micropillars showed a high yield strength (~ 200 MPa) but plastic deformation progressed without any hardening and with continuous strain bursts.

Secondary electron SEM micrographs of one micropillar of grain B are shown before (Fig. 3b) and after compression (Figs. 3c and 3d), the latter from two adjacent lateral sides (front side and left side). The marks on the top surface of the pillar before deformation (Fig. 3b) were created by the ion milling and did not influence the mechanical tests. Slip traces are clearly seen in the lateral surfaces of the micropillars. They were associated with (0001)<1-210> basal slip, as indicated by the schematic of the crystallographic lattice in Figs. 3e and 3f, although the highest SF for this orientation is attained for <c+a> pyramidal dislocations. The activation of basal slip is not surprising taking into account that the CRSS for basal slip is much lower than that for pyramidal slip and the large misorientation angle (17º) with respect to the *c* axis. Moreover, the low value of the yield stress -as compared with similar tests on other Mg alloys- seems to indicate that the initial yield was controlled by basal slip. In fact, the CRSS for basal slip, calculated from yield strength (marked by the black stars in Fig. 3a) and the corresponding SF of micropillars, was 14±3 MPa and this value is very close to the one determined from micropillars from grain A. Alizadeh et al. [40] found that micropillars of a Mg-4Zn (wt.%) deformed in compression along the *c* axis accommodated the plastic deformation by basal slip and they did not show any strain hardening. This result is in contrast with the strong hardening in Fig. 3c. That seems to point out that pyramidal slip was also active in the Mg-Ca-Zn alloy and was responsible for the strain hardening, in agreement with the behavior found in Mg-Al [42] and Mg-Y [24] alloys.

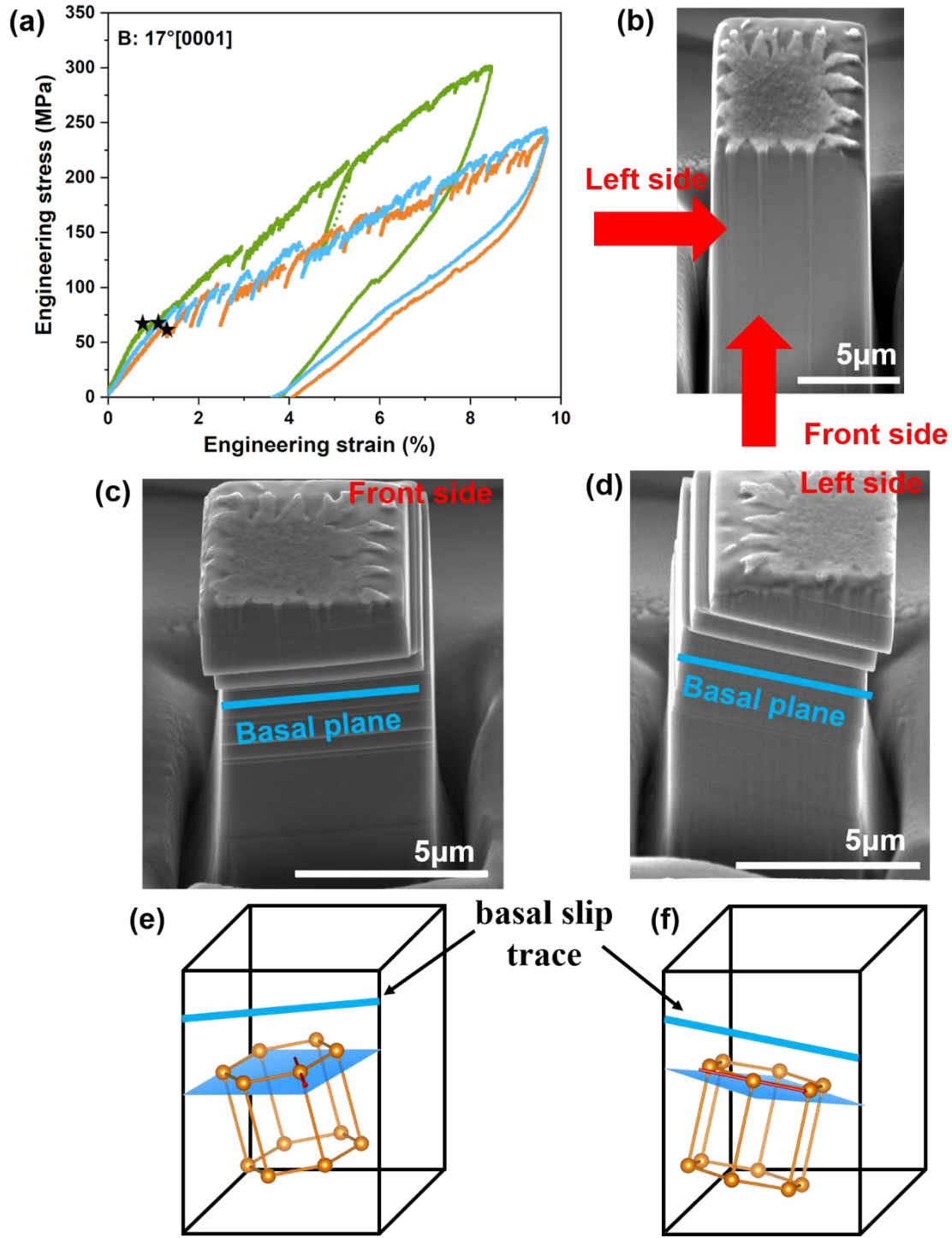

**Fig. 3.** (a) Representative engineering stress-strain curves of micropillar compression tests in micropillars carved from grain B. The black starts indicate the yield point in each curve. (b) SEM micrograph of the micropillar before deformation and after compression observed from (c) the front side and (d) the right side. The corresponding schematics of the crystallographic lattice in the outline of the pillar are shown in (e) for front side and (f) for right side. The blue planes indicate the basal glide planes, and the red arrows represent the relevant shear directions.

To clarify these points, compression tests of micropillars from grain B were carried out *in situ* within the SEM to ascertain the evolution of slip traces. The video of the test and the corresponding load-displacement curve can be found in the Supplementary Material (video V1), which shows that the apparition of basal slip traces is associated with strain bursts when the applied strain is around 2%. Further apparition of basal slip bands in the lateral surface of the micropillar was always associated with strain bursts, characterized by a sudden reduction of the load. Thus, it can be concluded that basal slip was activated during the deformation of the micropillar but the activation of pyramidal slip cannot be ruled out.

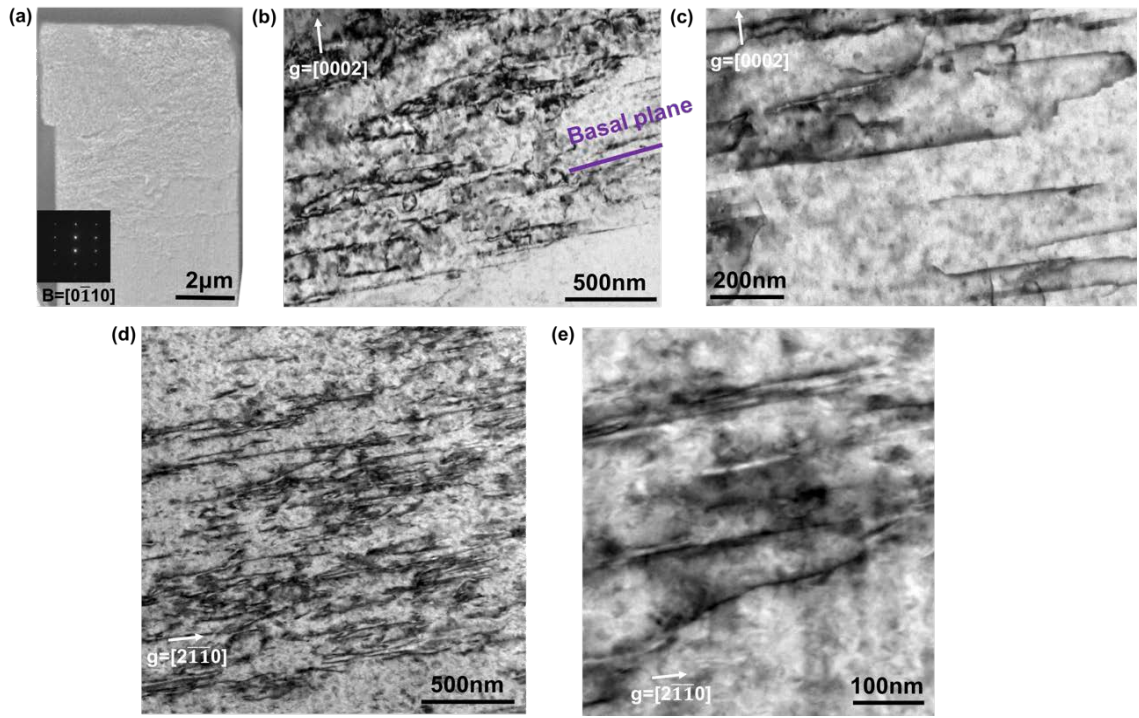

**Fig. 4.** (a) Low magnification bright field TEM micrographs of the lamella extracted from a deformed micropillar carved from grain B. The diffraction pattern in the inset shows that the lamella was parallel to the $(0\bar{1}10)$ plane. (b) and (c) Bright field TEM micrographs under the two-beam condition with g=[0002], revealing the dislocations containing <c>-component; (d) and (e) bright field TEM micrographs showing <c+a> dislocations in the same area under g=$[2\bar{1}\bar{1}0]$ condition.

The dislocation structure in a micropillar compressed up to 10% was analyzed by TEM in the thin lamella parallel to the $(0\bar{1}10)$ plane that was extracted from one of the deformed micropillars (Fig. 4a). It is obvious that twins were not nucleated during deformation. Bright field TEM micrographs at higher magnification with g=[0002] revealed the presence of a large density of dislocations with <c> component (Figs. 4b and 4c). They also appear in the same area with g=[2$\bar{1}\bar{1}$0] (Figs. 4d and 4e), indicating that they are pyramidal <c+a> dislocations. The straight dislocations parallel to the basal plane (as indicated by the blue line in Fig. 4b) result from the straight edge or near edge part in <c+a> pyramidal I or pyramidal II slip planes [24]. Afterwards, the thin foil was tilted and observed within the (1$\bar{2}$10) plane. The dislocation structures revealed under g=[0002] and g=[10$\bar{1}$1] are shown in Figs. 5a and Fig. 5b, respectively. The dislocation density is large under g=[0002] (Fig. 5a), while most of the dislocations are out of contrast when g=[10$\bar{1}$1] within the same region, indicating that <c+a> dislocations contain the 1/3[$\bar{1}\bar{1}$23] Burgers vector. Thus, the strong work hardening region was due to the activation of the pyramidal <c+a> dislocations.

Further indirect evidence of the activation of pyramidal slip is given by the apparent Bauschinger effect upon unloading (Fig. 3a). This phenomenon can be attributed to the accumulation of <c+a> dislocations in the micropillar (Fig. 4a) that facilitate reverse dislocation slip upon unloading. It should be noted that the Bauschinger effect is less pronounced in the micropillars deformed by basal slip (Fig. 2) because basal dislocations are not stored within the micropillar but leave the micropillar and lead to slip steps on the surface (Figs 2c and 2d). On the contrary, slip traces and slip steps associated with pyramidal <c+a> dislocations were not found on the surfaces on the micropillars in Fig. 3.

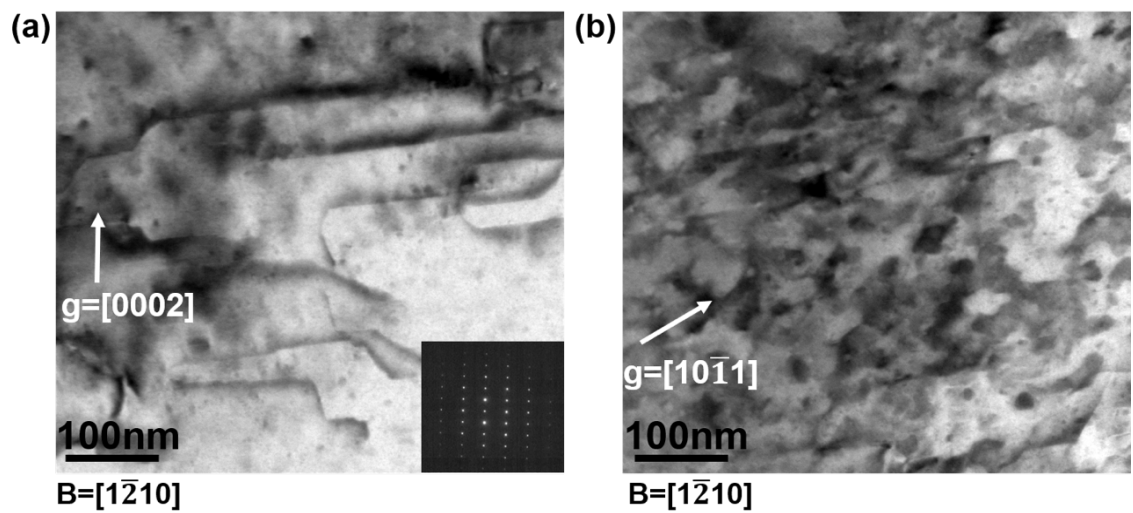

**Fig. 5.** Bright field TEM micrographs observed of the lamella extracted from a deformed micropillar carved from grain B. The observation plane was parallel to (1$\bar{2}$10), as indicated by the inset diffraction pattern, under two beam conditions: (a) g=[0002] and (b) g=[10$\bar{1}$1].

In order to ascertain the onset of pyramidal slip, one micropillar extracted from grain B was deformed up to the engineering stress of 175 MPa and unloaded. A SEM micrograph of the deformed micropillar is shown in Fig. 6a, and the slip traces along the basal plane (marked by the dashed blue line) are clearly seen. In order to elucidate the deformation mechanism, a thin foil parallel to the (0$\bar{1}$10) plane was lift out from the deformed micropillar and observed by TEM (Fig. 6b). The corresponding diffraction pattern in Fig. 6b indicates that twining did not occur. The dislocations structures within the deformed region were analyzed by TEM under g=[0002] (Fig. 6c) and g=[2$\bar{1}\bar{1}$0] (Fig. 6d) conditions and some dislocations were visible, indicating that activation of pyramidal <c+a> slip took place at stresses lower than 175 MPa and it was responsible for the strong hardening observed in these micropillars after the onset of yielding. On the basis of this evidence, the CRSS for <c+a> pyramidal slip has to be ≤ 85 MPa taking into account the SF for pyramidal slip (Table 1).

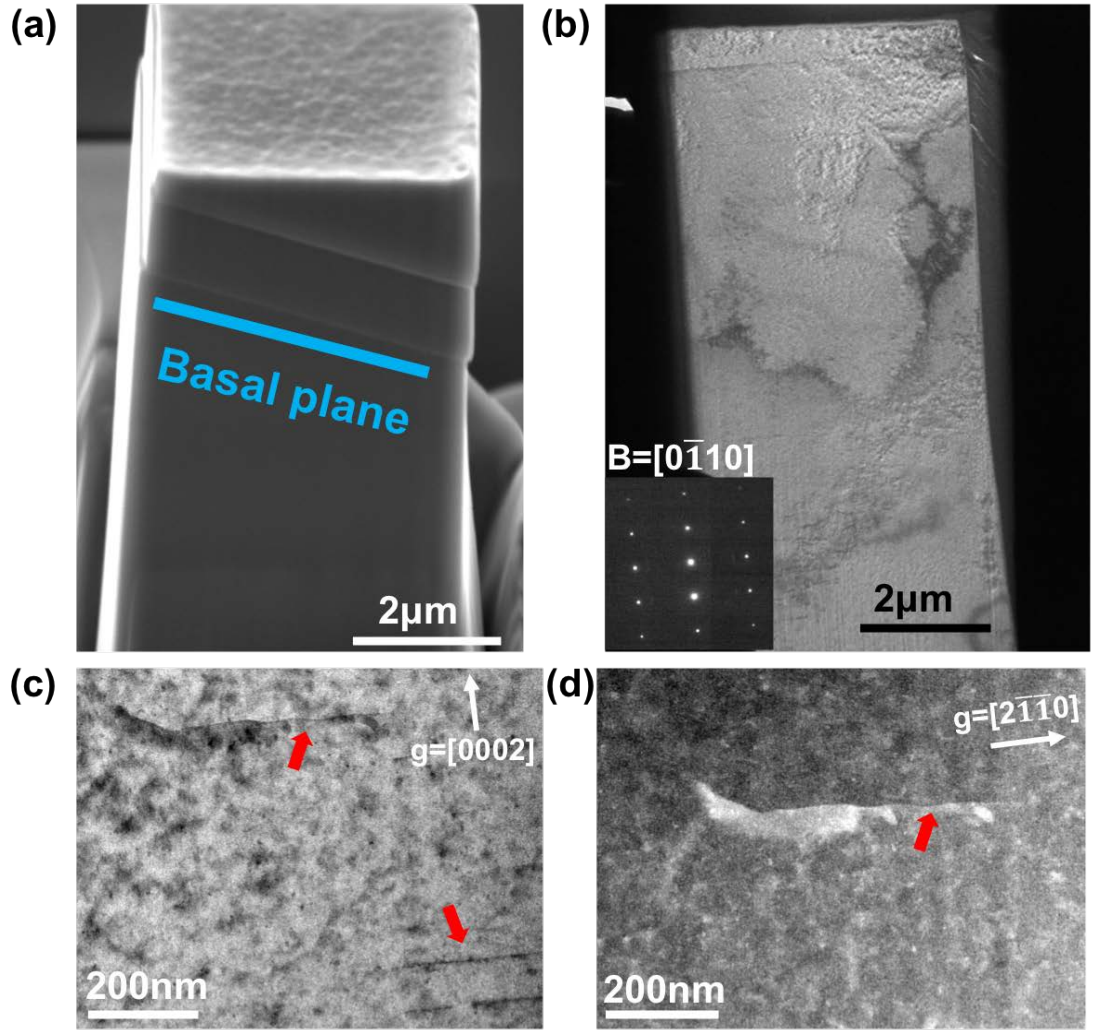

**Fig. 6.** (a) SEM micrograph of a micropillar carved from grain B and deformed up to an applied stress of 175 MPa. Slip traces parallel to the basal plane (marked by dashed blue line) are clearly visible. (b) Bright field TEM micrograph of the thin lamella parallel to the $(0\bar{1}10)$ plane extracted from the micropillar. (c) Bright field TEM micrograph under the two-beam condition with g=[0002] and (d) Dark field TEM micrograph under g=[2$\bar{1}\bar{1}$0]. Dislocations are indicated by the red arrows.

### 3.4 Deformation mechanisms of micropillars in grain C (nearly [11-20])

The engineering stress-strain curves of the micropillars deformed along [11-20] are plotted in Fig. 7a. Two curves are separated by shifting the x axis by 1% for the sake of clarity. Distinct strain bursts appeared after the initial elastic region (marked with red stars) and the stress dropped suddenly. The deformation continued when the stress reached the points marked with black triangles and the stress carried by the micropillar increased with strain up to 4% and deformation was accompanied by smaller strain bursts. Further deformation occurs at constant stress although there are many strain bursts as deformation progresses. The critical stress at the strain burst (denoted by the red star) was 78±33 MPa and the stress after the strain burst (denoted by the black triangles) was 63±2

MPa. The large scatter in the curves one can be ascribed to the misalignment of the contact between the indenter and the micropillar surface because the twin nucleation stress is very sensitive to the local stress state. However, this initial scatter did not influence the deformation of the micropillars afterwards.

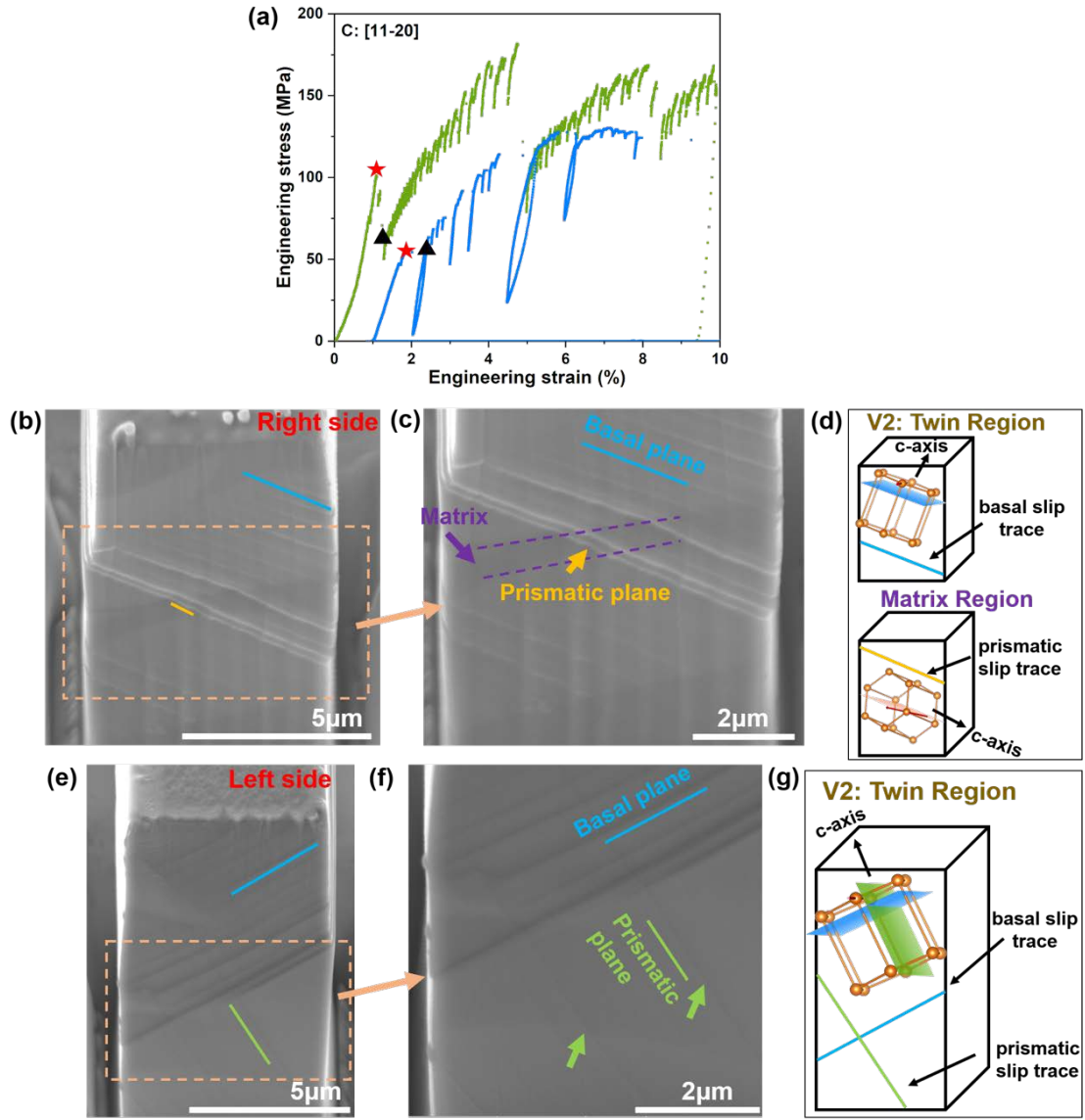

**Fig. 7.** (a) Representative engineering stress-strain curves of micropillar compression tests in micropillars carved from grain C. The red stars indicate the critical stress for twin nucleation and the black triangles for twin growth. (b) SEM micrograph of the micropillar observed from the right side and (c) enlarged region as marked in (b). (d) Schematics of the crystallographic lattice in the matrix and in the twin regions located in the outline of the pillar. (e) SEM micrograph of the micropillar observed from the left side, and (f) enlarged region of (e). (g) Schematics of the crystallographic lattice in the twin regions located in the outline of the pillar. The blue and green lines correspond to the traces of basal and prismatic glide planes in the twin region, while the orange line stands for the trace of the prismatic glide plane in the matrix region.

SEM images of the left and sides of the deformed micropillar C are depicted in Figs. 7b and 7e, respectively. Two regions with different contrast are revealed in Fig. 7b, as well as massive slip traces, marked by blue and orange dashed lines. The middle region of the deformed pillar was enlarged and is shown in Fig. 7c for the sake of clarity. A small region with dark contrast, marked by purple dashed lines, can be identified in the middle of the micropillar and it is surrounded by bright regions. Many slip traces appear in the bright region parallel to the blue line, and cross into the other region, as shown by the orange line. The morphology of the as-compressed micropillar observed from the left view side is depicted in Figure 7e. Massive slip traces are still visible along the lateral surface of the deformed pillar and distinct slip traces -perpendicular to each other and marked by blue and green lines- are distributed along the lateral surface (Fig. 7f). In summary, several features appear in the different lateral surfaces, including two regions with different contrast, slip transmission through these regions and straight perpendicular slip bands.

A thin lamella was extracted from the deformed micropillar along the right view side, as shown in Fig. 7b, and characterized by TKD. The TKD map in Fig. 8 reveals that two twin variants have appeared during the deformation. The smaller one, V1, was located at the upper left corner (light green color) and the larger one, V2, occupied the most part of the micropillar (green color). The V1 and V2 twin variants were identified as {0-112}<01-11> and {1-102}<-1101> based on the TKD result, and the latter presents the highest SF among the twin variants. Even though the highest SF in the parent crystal corresponded to the <c+a> pyramidal II slip system, tensile twining dominated the deformation because the lower CRSS for twin nucleation. The distribution of twin variants and matrix are consistent with the bright and dark contrast regions on the lateral surfaces in Fig. 7b. It is apparent that the slip traces in Fig. 7b are associated with the V2 twin region. The difference of the contrast between two regions are ascribed to the lattice distortion introduced by twin nucleation.

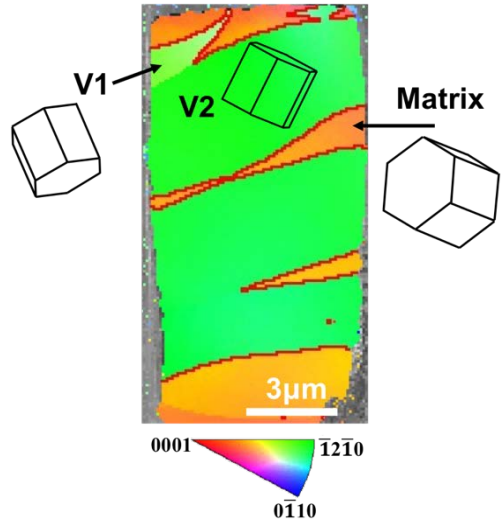

**Fig. 8**. TKD map of the lamella extracted from the deformed micropillar of grain C, showing the parent matrix (orange) and the twin variants V1 (light green) and V2 (green). The respective lattice orientations are also shown.

The crystallographic planes and corresponding slip directions are plotted along the two lateral views including right side (Figs. 7b-7c) and left side (Figs. 7e-7f), based on the orientation of the V2 twin variant and of the matrix. The slip traces within the twin region in Fig. 7c are parallel to the basal plane, as denoted by the blue lines, whilst those formed in the matrix belong to the prismatic plane, as marked by the orange line. It confirms that (0001)<1-210> basal slip was active in the twin region, while (1-100)<11-20> prismatic slip developed in the matrix region because they presented the highest SFs. According to the schematics of the crystallographic lattice in Fig. 7g, the slip traces perpendicular to each other within the twin region in Fig. 7f are associated with basal slip and prismatic slip, as denoted by the blue and green lines, respectively. Activation of prismatic slip in both matrix and twin regions is rarely found in Mg alloys with the exception of Mg-Y-Zn alloys [54]. These results show that the addition of Zn and Ca can promote the activation of prismatic slip within the twinned regions.

The co-existence of several deformation modes during plastic deformation makes difficult to identify which one leads to yielding. However, the first strain burst after the elastic region in Fig. 7a (marked by the red stars) can be ascribed to the nucleation of the twin, while the critical point after the strain burst (marked by the black triangles) is associated with the twin growth, as demonstrated previously in pure Mg [37] and Mg-Al alloys [42]. Therefore, the measured CRSS for tensile twin nucleation and twin growth were 38±16 MPa and 31±1 MPa, respectively, in Mg-Ca-Zn alloys. In addition, prismatic slip activity were detected during deformation but it is not possible to determine the CRSS for prismatic slip. However, it can be noticed that prismatic slip took place when the applied stress was ≤ 175MPa.

In order to ascertain the dislocation structures, a thin lamella extracted from the micropillars was analyzed by TEM. The distribution of matrix and twin regions in the lamella is shown in the low magnification TEM micrograph in Fig. 9a. The corresponding twin boundaries are indicated by the pink arrow. The twin region in the upper part with B=[$\bar{1}\bar{1}20$] was analyzed at higher magnification and observed under the two beam conditions. The corresponding dark field and bright field images with the diffraction vector g=[0002] are depicted in Figs. 9b and 9c. All the visible dislocations in both micrographs contain a Burgers vector with <c> component. Some long dislocations are lying on the basal plane, as well as some black dots, highlighted by blue lines. Since basal planes are edge-on to B, the visible dislocations under g=[0002] are the projections of <c+a> dislocations in pyramidal planes. Some of dislocations are still in contrast when the foil is tilted to g=[1$\bar{1}$00] diffraction, as indicated in Fig. 9d, suggesting the existence of <c+a> dislocations. The activation of <c+a> dislocations on both pyramidal slip I and II planes was also found in Mg-Y [24,55] alloys and it was associated with the reduction of the CRSS ratio between basal and non-basal slip due to the presence of solute atoms. In addition, the activation of <c+a> pyramidal slip system is consistent with the high SF of pyramidal slip within the twin region (Table 3). Assuming that the CRSS for pyramidal slip is 85 MPa and taking into account the SF for pyramidal <c+a> slip, compressive stresses of 172 MPa are necessary to promote the motion of <c+a> dislocations, which were attained in one of the tests (Fig. 7a). In addition, some long and straight dislocations are also found on the prismatic plane, as indicated by the green lines in Figs. 9b and 9c. They are out of contrast when g=[1$\bar{1}$00], indicating that they are pure <c> dislocations. A large density of <c> dislocations in the twin region has been reported previously [18,36,56–58] and they may result from the dissociation of <c+a> dislocations [11,18] or from the transformation of <a> dislocations in the matrix after twinning [36].

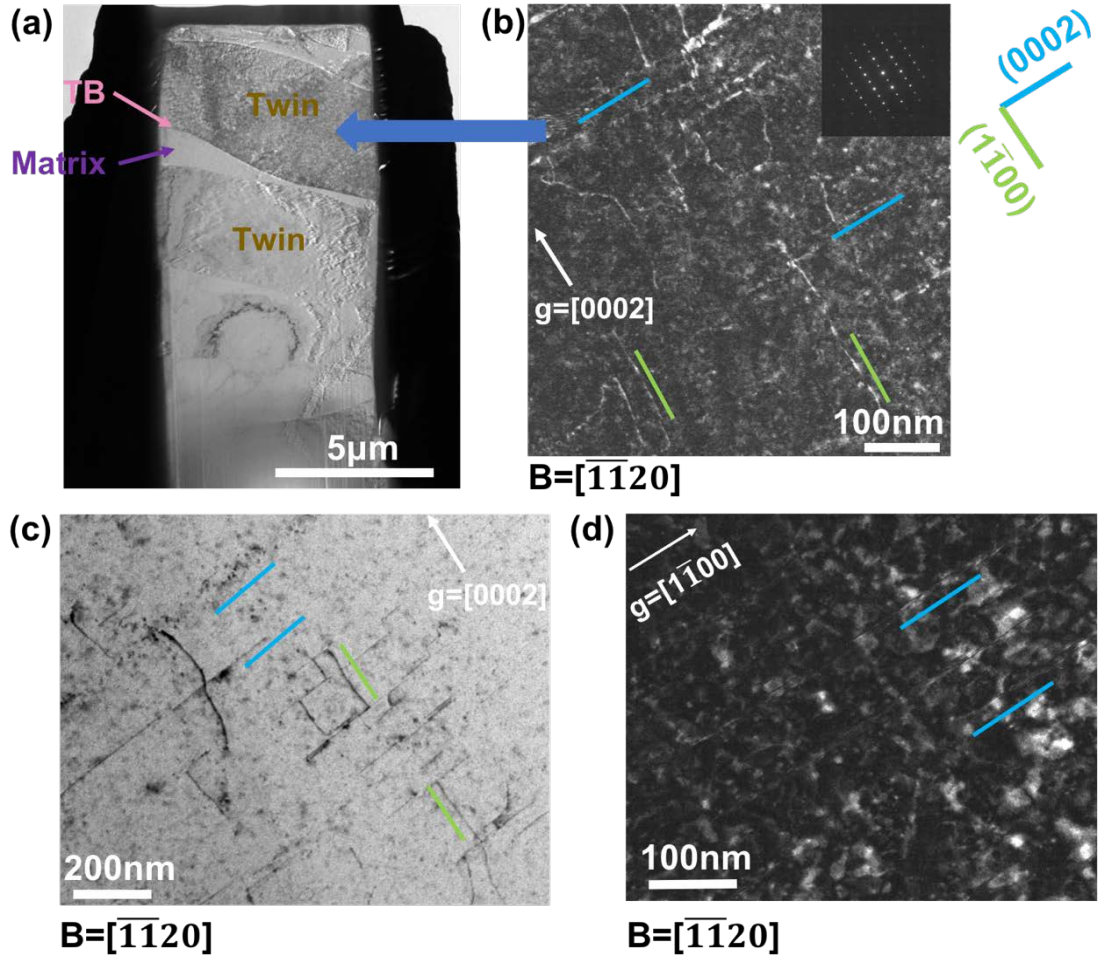

**Fig. 9**. TEM micrographs of a thin lamella extracted from the deformed micropillar carved from grain C. (a) Low magnification bright-field micrograph showing the twin and matrix regions. (b) Dark field micrograph and (c) bright field micrograph of dislocations in the twin region under g=[0002]. The trace of basal plane is given by the blue line and the trace of the prismatic plane by the green line. (d) Dark field micrograph showing dislocations in the twin region under g=[1$\bar{1}$00].

**Table 3.** Highest SFs among the individual deformation modes within the twin regions after compression of micropillars from grains C and D.

| Initial loading direction | Inclination angle within the twin region | Highest Schmid Factors | | | | | |
|---|---|---|---|---|---|---|---|
| | | Basal slip | Tensile twin | Prismatic slip | Pyr I \<a\> | Pyr I \<a+c\> | Pyr II \<c+a\> |
| C [11-20] | 27° | 0.35 | 0.38 | 0.10 | 0.25 | 0.47 | 0.45 |
| D [11-10] | 9° | 0.14 | -* | 0.01 | 0.06 | 0.47 | 0.49 |

*: Tensile twin cannot be activated during compression along the *c* axis.

## 3.5 Deformation behavior of micropillars in grain D (nearly [10-10])

Representative engineering stress-strain curves from the compressed micropillars carved from grain D ([10-10] direction) are plotted in Fig. 10a. For the sake of clarity, the curves are shifted by 2% along the x axis. The initial elastic region is followed by strain burst marked by red stars. After the stress drop, deformation resumed when the stress reached the value marked by the black triangle and the stress-strain curves show continuous hardening which is associated with small, continuous strain bursts. The critical stresses before and after the strain bursts are determined to be 76±21 MPa and 88±9 MPa, as marked by the red stars and black triangles, respectively. These curves are different from those obtained in the same orientation from micropillars of pure Mg [34,36,37,39], Mg-Al [41], Mg-Zn [39] and Mg-Y [24]. In these materials, the initial strain burst was followed by a stress plateau and strong hardening, and these regions in the stress-strain curves were associated with twin nucleation, twin growth and activation of other slip systems.

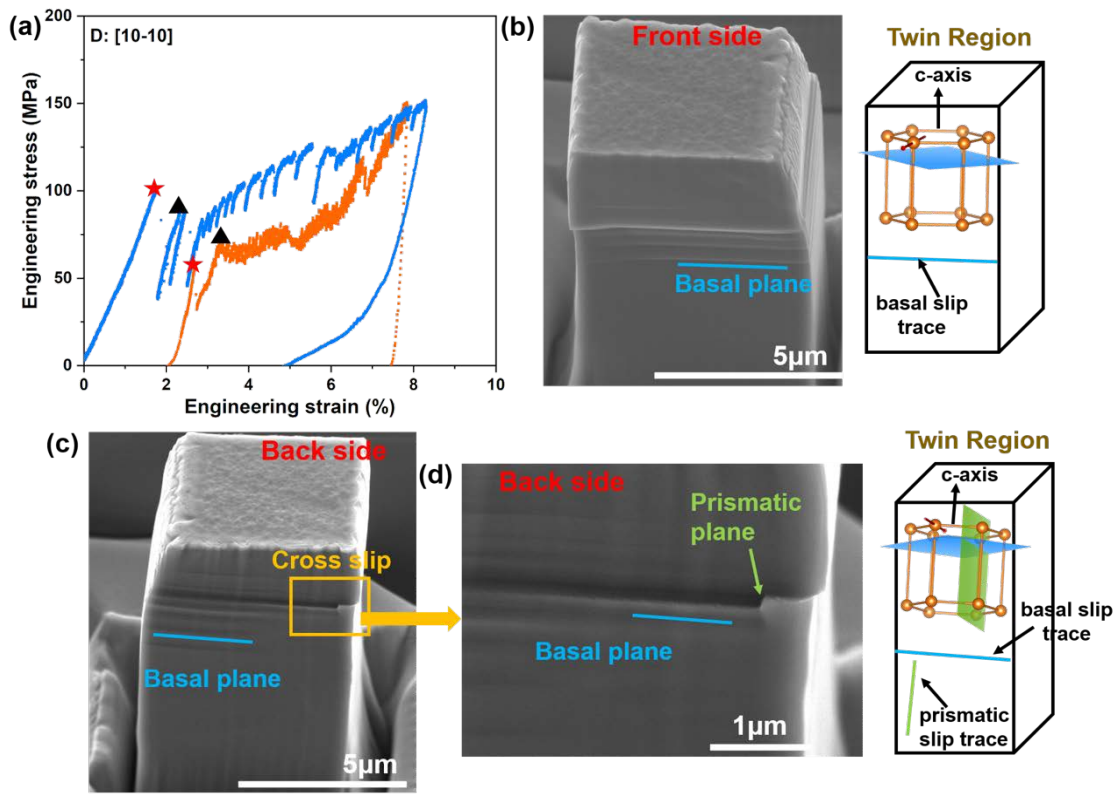

**Fig. 10.** (a) Representative engineering stress-strain curves of micropillar compression tests in micropillars carved from grain D. The critical stress for twin nucleation in marked by red stars and the critical stress for twin growth by black triangles. (b) SEM micrograph of the micropillar observed from the front side. (c) SEM micrograph of the micropillar observed from the back side. (d) Detail of (c) showing cross-slip between basal and prismatic planes. The blue and green lines correspond to the traces of basal and prismatic glide planes in the twin region.

The morphology of the front and back sides of compressed micropillar carved from grain D is shown in the SEM micrographs of Figs. 10b and 10c, respectively. Horizontal slip traces are revealed within the deformed pillars, close to the blue line. They are also found on the back side (Fig. 10d) as well as traces of cross-slip that can be observed at higher magnification in Fig. 10d. To further confirm the deformation mechanisms, a thin foil was extracted parallel to the front lateral side in Fig. 10c, and analyzed by TKD. Most of the micropillar was twinned (blue color in Fig. 11), with the exception of several small regions at the top and at the center of the pillar (light green color in Fig. 11). Activation of twinning is not surprising in this orientation and deformation progresses afterwards by plastic slip in the twin region, as it was also reported in micropillars of Mg-Al alloy deformed along similar orientation [41]. However, the presence of prismatic slip traces and of cross-slip between basal and prismatic planes in the twin region has not been reported in other Mg alloys.

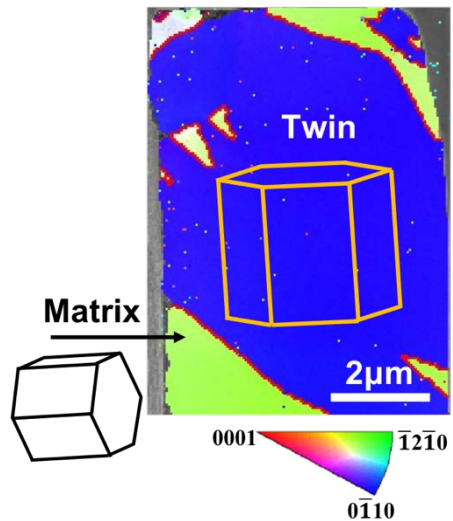

**Fig. 11**. TKD map of the lamella extracted from the deformed micropillar of grain D, showing the parent matrix (green) and the twin region (dark blue). The respective lattice orientations are also shown.

The morphology of the deformed micropillar and the deformation mechanisms of the micropillar carved from grain D (twinning and basal slip in the twin region) are qualitatively similar to those reported in micropillars of pure Mg and Mg-Al alloys with the same orientation [41]. However, there are large quantitative differences in the stress-strain curves. The CRSS for twin nucleation and twin growth in micropillars carved from grain D are 37±10 MPa and 42±4 MPa, taking into account the applied stress es at the points marked by red stars and black triangles (Fig. 10a) and the corresponding SFs (Table 1). They comparable with those obtained in the case of micropillar C. They are very different for those measured in Mg-9Al (at.%) alloys [41] where the CRSS for twin nucleation was ~138 MPa and ~86 MPa for twin growth. The high CRSS for twin growth and the marked strong hardening suggest the presence of Ca and Zn hinders twin boundary migration.

## 4. Discussion
### 4.1 Effect of Zn and Ca on the CRSSs

Mechanical tests at the micron or sub-micron scale usually lead to an overestimation of the yield strength as compared with the bulk properties [59–68]. This "size effect" is related to the limited number of mobile dislocations and/or to the scarcity of dislocation sources in the small volume, and the critical size of the micropillar to determine the bulk properties depend on the particular alloy and microstructure. For instance, Liu et al. [34] reported that micropillars of 10 μm in diameter are necessary to measure the CRSS for basal slip of the bulk in pure Mg. On the contrary, the size effect was negligible for micropillars of > 3.5 μm in diameter in Mg-10Al (wt.%) alloys [59]. In addition, negligible size effects were reported in a previous investigation to determine the CRSS for different slip systems and twinning in Mg-Al and Mg-Zn alloys by means of compression of square micropillars of 5x5 μm$^2$ cross-section [41,42]. Thus, it can be assumed that the size effect on the yield stress is small in the Mg-Ca-Zn micropillars tested in this investigation but it is very difficult to eliminate completely this effect. Nevertheless, the experimental conditions were the same for all the deformation modes and it is expected that similar size effects -if any- would be active in all cases. Thus, the effect of alloying on the CRSS for each deformation mode can be captured.

The yield stress in the stress-strain curves if the micropillars loading along different orientations is depicted in Table 4. The corresponding CRSS is also included in this Table from the SF for the dominant deformation mode at the onset of plasticity which was determined from the slip traces on the micropillar surface and the TEM analysis. Basal slip was found in the micropillars deformed along A and B directions while twin nucleation and growth were observed in the micropillars deformed along C and D directions. The differences in the CRSS for twin growth between grains C and D can be attributed to the misalignment contact during reloading.

**Table 4** Micro-compression yielding stress and critical resolved shear stress determined from micropillar compressed along every loading direction.

| Loading directions | Yield stress (MPa) | Critical resolved shear stress (relevant slip system) (MPa) |
|---|---|---|
| A [25-7-6] | 27±1 | 13±0.5 (basal slip) |
| B [0001] | 62±3 | 14±3 (basal slip) |
| C [11-20] | 78±33 | 38±16 (twin nucleation) |
| C [11-20] | 63±2 | 31±1 (twin growth) |
| D [11-10] | 76±21 | 37±10 (twin nucleation) |
| D [11-10] | 88±9 | 42±4 (twin growth) |

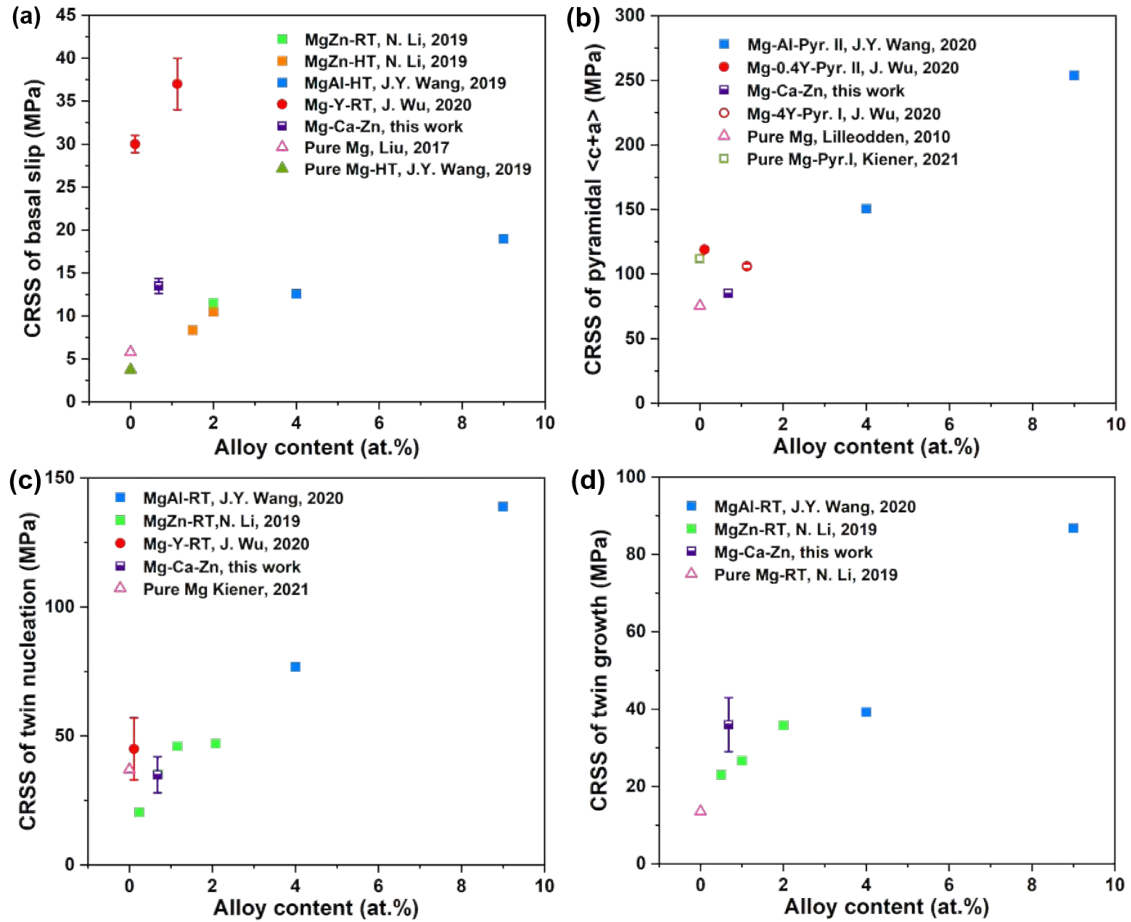

**Fig. 12**. CRSS of different deformation modes in Mg and Mg alloys measured from compression of micropillars of around 5 μm. The results of Mg-Ca-Zn alloys combined with pure Mg [33–35,60] and other Mg alloys are included [24,41,42,60]. (a) <a> basal slip; (b) <c+a> pyramidal slip, (c) twin nucleation and (d) twin growth.

The CRSS for basal slip, twin nucleation, twin growth and pyramidal <c+a> in Mg-Ca-Zn alloys are plotted in Fig.12 together with those obtained in pure Mg [34,35,60] and other Mg alloys [24,41,42,60]. All the CRSS values were measured in micropillars of around 5 μm. It is obvious that the addition of Ca and Zn significantly improves the strength of basal slip compared to pure Mg [34]. Furthermore the co-addition of Ca and Zn behaves more effective to increase the CRSS for basal slip than just Zn or Al [42], but Y in solid solution is even more effective [24] (Fig. 12a). The large strengthening effect of Ca has been attributed to the shear modulus mismatch with the Mg matrix [61] and to the segregation of Ca atoms around basal dislocations [27].

The addition of Ca and Zn has limited strengthening effect on pyramidal slip, as compared to pure Mg [35] and Mg-Y [24] (Fig. 12b), indicating that <c+a> pyramidal slip is preferentially activated in Mg-Ca-Zn alloys. The CRSS for twin nucleation in Mg-Ca-Zn alloys is similar to the one found in Mg-Al [41] and Mg-Zn [60] and slightly lower than Mg-Y [24] (Fig. 12c). However, a high CRSS is required for twin growth in Mg-Ca-Zn alloys as compared with pure Mg [60], Mg-Al [41] and Mg-Zn [60] (Fig. 12d).

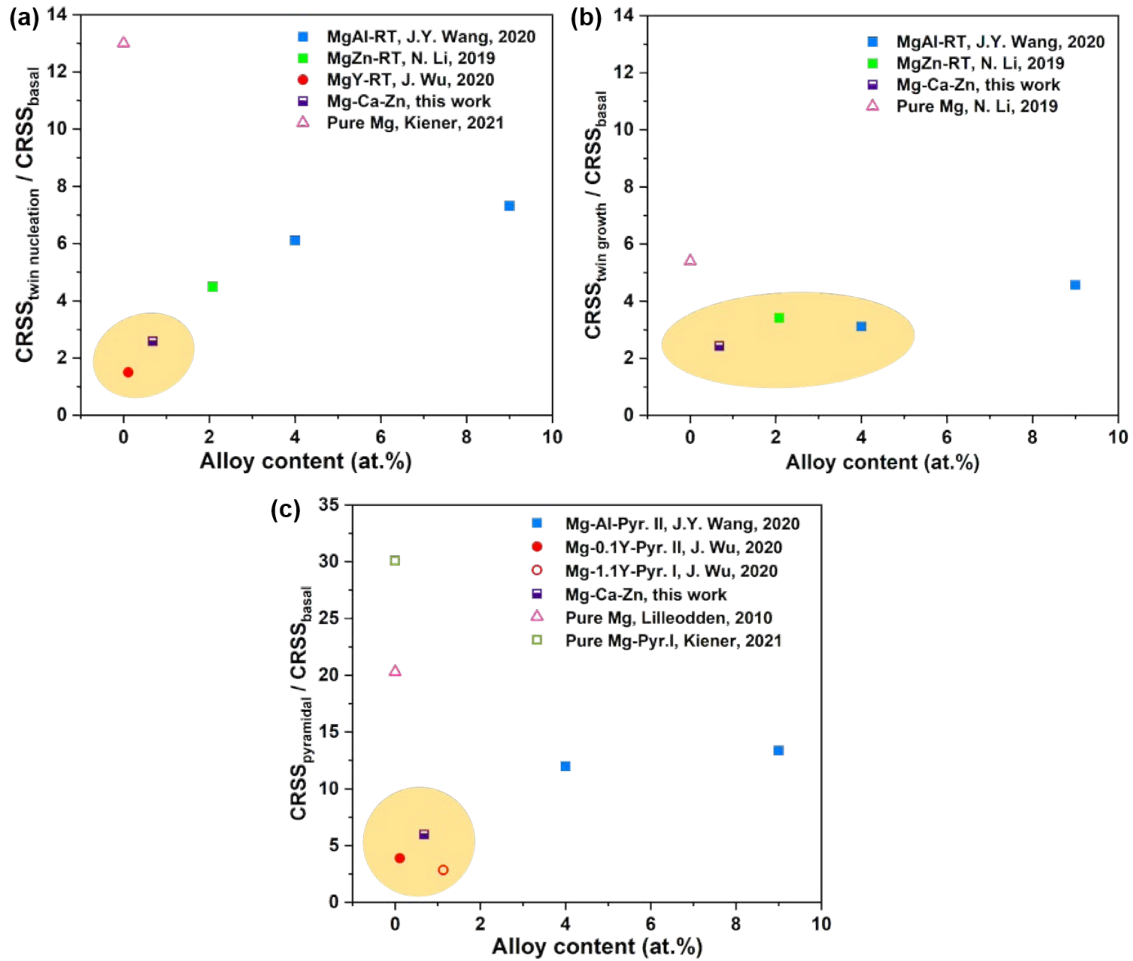
24

**Fig. 13**. (a) Ratio between the CRSS for twin nucleation and basal slip in pure Mg [33–35,60] and Mg alloys as a function of alloy content [24,41,42,60]. (b) *Idem* for twin growth and basal slip. (b) *Idem* for pyramidal slip and basal slip.

The plastic anisotropy of Mg alloys is dictated by the ratio between the CRSS for non-basal deformation modes (twin nucleation, twin growth and pyramidal slip) and basal slip. These ratios are plotted in Figs. 13a, b and c, respectively, for the Mg alloys in Fig. 12. The results in Fig. 13 show that pure Mg presents very high plastic anisotropy and but the addition of Ca and Zn dramatically reduces the $CRSS_{twin\ nucleation}/CRSS_{basal}$ and $CRSS_{pyramidal}/CRSS_{basal}$ ratios. Thus, Mg-Zn-Ca and Mg-Y alloys are prone to present the best formability. Moreover, the evidence of prismatic slip indicated that the ratio $CRSS_{prismatic}/CRSS_{basal}$ was also reduced in Mg-Ca-Zn alloys and this phenomenon can also help to improve the ductility of this alloy [8].

### 4.2 Effect of Ca and Zn on twin growth

The results of micropillar compression tests normal to the *c* axis in grains C and D (along [11-20] and [10-10]) of Mg-Ca-Zn alloys showed that the CRSS for twin growth was higher than that found in either Mg-Zn and Mg-Al alloys (Fig. 12d). Moreover, twin growth occurred at constant stress in Mg-Al [41] and Mg-Zn [60] alloys, but it led to strain hardening in the present Mg-Ca-Zn alloys (Figs. 7a and 10a). While solid solution strengthening for twin nucleation has been found in several Mg alloys [24,41,60,62], solid solution strengthening of twin growth has not been reported.

It has been hypothesized by both experimental results [63–68] and theoretical analyses [69–71] that segregation of solute atoms could stabilize the twin boundary and increase the CRSS for twin growth. In addition, twin growth can be hindered by the solute strengthening effect on twinning dislocations [62]. The latter effect increases with the differences in size and electronegativity between Mg and solute atoms. The atomic radius of Mg is 0.16nm, much smaller than that of Ca (0.197nm [71]), while the electronegativity of Mg is 1.2 and that of Ca 1.0. Thus, the size misfit between Ca and Mg atoms can also hinder the mobility of twinning dislocations and contribute to increase the CRSS for twin growth. Moreover, the reduction of twin boundary migration can influence the texture evolution and effectively weaken the basal texture caused by twinning during the thermomechanical processing [72]. This mechanism is responsible for the random texture of Mg-Ca-Zn alloys and helps to reduce the plastic anisotropy and improve the ductility of these alloys [16].

### 4.3 Effect of Ca and Zn on the activation of prismatic slip

Slip traces lying on the prismatic plane were observed within the twin and matrix regions of micropillars deformed along the *a* axis ([11-20] and [10-10]) (Figs. 7c and 10c). It was not possible, however, to determine the CRSS for prismatic slip since micropillar yielding was dominated by twinning and simultaneous activation of basal and prismatic slip occurred at later stages during deformation. Nevertheless, prismatic slip occurred in the micropillars in Fig. 7a and the maximum stress applied to these micropillars was 175 MPa. However, no prismatic slip was found in micropillars of solution-treated Mg-5Zn (wt.%) alloys when the maximum stress reached 300 MPa [39]. Thus, it can be concluded that the addition of Ca reduces the CRSS for prismatic slip. This result was predicted by density functional theory calculations [30] which showed that Ca can significantly reduce the stacking fault energy on prismatic planes and soften the prismatic slip. This reduction in the stacking fault energy due to the presence of Ca was maintained in Mg-Ca-Zn alloys [32]. Of course, activation of prismatic slip favored the apparition of basal to prismatic cross-slip, which was observed on the lateral surface of deformed micropillars (Fig. 10d). The activation of prismatic slip as well as of basal to prismatic cross slip is consistent with the excellent ductility and formability of Mg-1.8Zn-0.2Ca [19] and Mg-1.5Zn-0.1Ca alloys [73], both of which were attributed to the presence of <a> prismatic slip.

### 4.4 Effect of Ca and Zn on the activation of pyramidal slip

Pyramidal <c+a> dislocations were found in micropillars compressed close to the *c* axis (grain B) and in the twin region of the micropillars loaded along the *a* axis (grains C and D). These results are in agreement with previous micropillar compression tests in Mg-Al [41] and Mg-Y [24] alloys as well as with mechanical tests of polycrystalline Mg-Ca-X alloys [16,20,26,74–76]. For instance, Pan et al. [26] observed a large number of <c+a> dislocations in a Mg-1.0Ca-1.0Al-0.2Zn-0.1Mn (wt.%) alloy after deformation and suggested that they significantly contributed to the enhanced ductility of the material. Zhu et al. [75] found that the ductility of Mg-6Al-1Ca (wt.%) alloy was improved through the activation of pyramidal II <c+a> dislocations while the <c+a> pyramidal I dislocations were found in Mg–1Al–0.1Ca (wt.%) alloys with lower alloying content [76]. <c+a> Pyramidal II dislocations were also identified in the dilute Mg-0.3Zn-0.1Ca (wt.%) alloy [16], while Zhu et al. [20] reported that plastic deformation mediated by pyramidal I <a> dislocations in Mg-0.47Ca (wt.%) alloy. Thus, activation of pyramidal I or pyramidal II slip depends on the solute atoms and content, which modify the stacking fault energies for both slip systems. First principles calculations of Ding et al. [29] showed that Ca reduced the stacking fault energy of <c+a> pyramidal I while Zn has a negligible effect and support our experimental observations in Mg-Ca-Zn micropillars. Finally, first principles calculations [10] found that Ca and Zn increased the energy barrier for cross-slip from <c+a> pyramidal I to <c+a> pyramidal II. This result explains the absence of cross-slip between both pyramidal slips systems in the present work.

## 5. Conclusions

The CRSS for basal slip, tensile twinning and pyramidal slip has been measured in a Mg-0.2Ca-1.8Zn (wt.%) alloy by means of compression tests in single-crystal micropillars with different orientations. The active deformation systems in each orientation were determined by means of the analysis of slip traces on the micropillar surface, transmission Kikuchi diffraction and transmission electron microscopy observations. The main conclusions of this research are the following:

• The presence of Zn and Ca in solid solution increased the CRSS for <a> basal slip to approximately 13.5 MPa, much larger than the one for pure Mg. The CRSSs for twin nucleation and growth were very similar (around 37 MPa), as opposed to most Mg alloys in which the CRSS for twin nucleation is much higher than that for twin growth. The high CRSS for twin growth was attributed to the segregation of solute atoms to the twin boundary and to the solute strengthening of twinning dislocations.

• Activation of <c+a> pyramidal slip was observed in micropillars suitably oriented and the CRSS for pyramidal slip (< 85 MPa) was comparable to that of pure Mg. Thus, the ratio between the CRSS of pyramidal and basal slip was low and comparable to that of Mg-Y alloys. Thus, Mg-Ca-Zn alloys presented low plastic anisotropy.

• Deformation by <a> prismatic slip was found and it was promoted by the reduction in the stacking fault energy of prismatic planes by the co-existence of Zn and Ca atoms. Moreover, cross-slip between basal and prismatic planes was also observed.

• The reduction of plastic anisotropy, the increase in the CRSS for twin growth (which weakens the basal texture caused by twinning during the thermomechanical processing) and the activation of multiple slip systems (basal, prismatic, pyramidal) and twinning, together with cross-slip, are responsible for the large ductility and formability of Mg-Ca-Zn alloys.

**Acknowledgements**

This work was supported by the National Natural Science Foundation of China [Grant Nos. 52001199 and 51825101]. J.Y. Wang acknowledges the China Postdoctoral Science Foundation (No. 2020M671113). JLL acknowledges the support from the European Research Council (ERC) under the European Union's Horizon 2020 research and innovation programme (Advanced Grant VIRMETAL, grant agreement No. 669141) and from the Spanish Ministry of Science (HexaGB project, reference RTI2018-098245).